\documentclass{ws-procs10x7}

\usepackage{myalias}
\usepackage{epsfig}
\newcommand{\tmop}[1]{\operatorname{#1}}
\def\rat {\varepsilon'/\varepsilon}

\def\beq {\begin{equation}}
\def\eeq{\end{equation}}
\def\bea {\begin{eqnarray}}
\def\eea {\end{eqnarray}}

\def\bc {\begin{center}}
\def\ec {\end{center}}
\begin{document}
\title{Higgs physics and CP violation
%\footnote{Talk given at ICHEP04, Beijing}
}
\author{Yue-Liang Wu}
\address{Institute of theoretical physics,CAS, Beijing, China}
%\affiliation{Institute of theoretical physics,CAS, Beijing, China}
\author{Yu-Feng Zhou }
%\affiliation{Institut f\"ur Physik, Dortmund University,\\
%D-44221 Dortmund, Germany}
\address{Institut f\"ur Physik, Dortmund University,\\
D-44221 Dortmund, Germany}

%\email[]{Talk given at ICHEP04, Beijing}
%\date{\today}
\twocolumn[\maketitle\abstract{
%\begin{abstract}
  The CP violation in $K$ and $B$ decays are discussed with the emphasize on the
  precise prediction for $\epsilon'/\epsilon$ and the new physics effects from a
  class of models with additional Higgs doublets and fermions.
  The contributions from standard model with two-Higgs-doublet (S2HDM) to $K$, $B$
  mixing and decays are briefly reviewed. The
  possible large effects on CP violation in $B\to \phi K$ is discussed in
  an extended standard model with both an additional Higgs doublet and fourth
  generation quarks (S2HDM4). We show that although the S2HDM
  %Standard Model with two-Higgs-doublet
  and the Standard model with fourth generation quarks
  {\em{alone}} are not likely to largely change the effective $\sin 2
  \beta$ from the decay of $B \rightarrow \phi K_S $, the S2HDM4 model can easily account for
  the possible large deviation  of $\sin 2 \beta$ without conflicting with
  other experimental constraints.
%\end{abstract}
}]
%\preprint{}
%\keywords{}
%\pacs{}
%\maketitle
%%%%%%%%%%%%%%%%%%%%%%%%%%%%%%%%%%%%%%%%%%%%%%%%%%%%%%%%%%%%%%%%%%%%%%%%%%%%%%%%
%%%%%%%%%%%%%%%%%%%%%%%%%%%%-Begin-%%%%%%%%%%%%%%%%%%%%%%%%%%%%%%%%%%%%%%%%%%%%%
%%%%%%%%%%%%%%%%%%%%%%%%%%%%%%%%%%%%%%%%%%%%%%%%%%%%%%%%%%%%%%%%%%%%%%%%%%%%%%%%
%Understanding the origins of CP violation in the standard
%model(SM) and beyond are the central topics in the present day
%particle physics.
%
The origin of CP violation with the SM and beyond is currently under
intense investigation.
In this talk we will focus  on the CP violations in hadronic $K$ and
$B$ meson decays.
%
%With the recent experimental progresses, the
%existence of direct CP violation in $K$ system and the mixing
%induced one in $B$ decays are now well established, which
%quantitatively agree the CKM mechanism.  This is a triumph of the
%SM.
%However, without a deeper understanding of the low energy
%dynamics of QCD, it is still difficult to precisely confront the
%SM predictions with the experimental data. A good example is
%evaluation of the direct CP violation parameter
%$\epsilon'/\epsilon$.
%On the other hand, there is a common belief
%that the SM is not a fundamental theory.  The new physics effects
%at electro-weak scale can not be excluded.  The most recent
%measurements from $B$-factories, although not conclusive, have
%already shown puzzling results and imply the new physics beyond
%the SM.

The decays $K\to\pi\pi$ are  described by a low energy effective
Hamiltonian with short and long distrance contributions
obsorbed into Wilson coefficients and matrix elements of local
operators respectively.
%eq.(1)
%\begin{align}
%H=\frac{G_F}{\sqrt{2}}\xi_u
%\left\{\sum_{i=1}^8 (z_i(q^2,\mu^2)+\tau y_i(q^2,\mu^2) )Q_i\right\}
%\end{align}
%with $z_i(q^2,\mu^2)$ and $y_i(q^2,\mu^2)$ being the Wilson
%coefficients and $\xi_q=V^*_{qs}V_{qd}$, $\tau=\xi_t/\xi_u$ .
%$Q_i$'s are 4-quark operators.  For the definition of the
%operators and other notations, see ref.\ \cite{Hambye:1998sm}
%which we closely follow.
Matrix elements for two of these
operators, $Q_6$ and $Q_8$, are most important for the evaluation
of direct CP violation parameter $\rat$.
%(2)+(3)
%\begin{eqnarray}
%Q_6 & = &-2\sum_{q=u,d,s}\bar{s}(1+\gamma_5) q \bar{q}(1-\gamma_5)d,\n\\
%Q_8 & = &-3\sum_{q=u,d,s} e_q\bar{s}(1+\gamma_5)q\bar{q}(1-\gamma_5)d,
%\end{eqnarray}
%where $e_q=(2/3,-1/3,-1/3)$.
%TheQCD corrections included in the Wilson coefficients represent
%the short distance terms computed in perturbative QCD.
The Wilson coefficients
depend on scal $\mu$ and to
next-to-leading order (NLO) corrections in a more complicated way
which depends also on the renormalization scheme.  
%The $\mu$-dependence
%in the coefficients is expected to be cancelled by the scale
%dependence of the matrix elements
%of the operators
%introduced
%through the upper cut-off in the integrals,
%and the running quark masses.
%
% add Yue-Liang 's work here
%The evaluation of long distance contribution to $\rat$ is a hard
%task.  %There was a long debate over whether $\rat$ is compatible
%with zero or can reach the sensitivity of the current experiments,
%which actually depends on how the long distance effects are
%handled.
The main issue is the evaluation of long distance contributions.
It was the first time pointed out in
Ref.\cite{Wu92} that with long distance chiral-loop
corrections to operator $\langle Q_6 \rangle$, $\rat$ could reach
$\mathcal{O}(10^{-3})$.
%its accuracy was also investigated in
%Ref.\cite{Heinrich:1992en}, which was later confirmed by the
%measurements some time ago.
%
For a consistent description to $K\to \pi\pi$, one needs
a  matching between short and long distance terms.

In the naive matching approaches \cite{naiveMatch}with the cut-off scale of $M$ for matrix
element being  identified with short distance scale $\mu$ at about $1$ GeV, a large
correction originates from rescattering of the pions, i.e., $K\to
\pi\pi\to\pi\pi$,
%\vskip -0.5 cm
%\begin{figure}[htb]
%\begin{center}
%\epsfig{file=contact.eps,width=7cm}
%%\epsfig{file=exchange.eps,width=7cm}
%\caption{Feynman diagram  for $K\to \pi \pi$ with strong final state  interactions.
%}\label{contact}
%\end{center}
%\end{figure}
\noindent
where the first step involves the weak operators $Q_6$
or $Q_8$ to $\mathcal{O}(p^2/N_c)$ and the second process is the
strong pion-pion
scattering. %as shown in Fig. \ref{contact}.
The large dependence of the cut-off resides on the contact
$\pi-\pi$ scattering which is known to have a bad high-energy behaviour
violating unitarity and needs to be moderated by some other amplitudes which
restore unitarity.
%
%\begin{figure}[htb]
%\begin{center}
%\epsfig{file=exchange.eps,width=7cm}
%\caption{Feynman diagram for $K\to \pi \pi$ with a vector-meson exchange.}\label{exchange}
%\end{center}
%\end{figure}
A standard prescription to restore unitarity is to introduce vector-meson
exchange diagrams.  For the $\pi\pi\to\pi\pi$ scattering one shall use the
contact and the $\rho$ exchange diagrams. It can be accomplished by using a
chiral Lagrangian for pseudo scalars with the introduction of vector mesons in
the lowest order\cite{Bando87}. The calculation of the one-loop diagrams
with a strong vertex was extended with the addition of a $\rho$-exchange
diagram.  The $\rho$ is included to symbolically represent the effects of all
other vector mesons (like $K^*$)
\cite{Kundu04}.

%Inaddition the pions are in $I=0$ or $I=2$ states and the exchange of
%$\rho$-mesons appears only in the $t$-channel.% see Fig.\ref{exchange}.
%%\vskip -1cm

In order to restore unitarity it is demanded that quadratic
divergences cancel between the contact and the $\rho$-exchange
diagrams.  It is indeed heartening to note that they come with
opposite signs, and cancel exactly if the following relation is
satisfied\cite{Kundu04}
%eq.(4)
\begin{align}\label{cancellation}
\frac{h^2}{m^2_{\rho}}=\frac{1}{3F_\pi^2}\, .
\end{align}
Here $h$ is the $\rho\pi\pi$ coupling strength and $F_\pi$ is the
pion decay constant ($F_\pi\approx 92 MeV$). This relation is to
be compared with the celebrated KSFR relation in which the factor is
2. The logarithmic
divergences still remain and should be matched to the QCD
logarithms in the region $m_{\rho}$ to $\mu \simeq 1$ GeV.

Though an exact matching to QCD needs to be checked, an improved
stability of the values for $\epsilon'/\epsilon$ was arrived in
the region from $m_{\rho}$ to $\mu \simeq 1$ GeV.
The main conclusion is that the presence of vector mesons improves
the calculation of the matrix elements by making them more stable
functions of the cut-off in the naive matching approach.
%The first results are encouraging to continue and complete the calculation. We
%are in the process of calculating the factorizable terms and presenting also the
%constant contributions.  We are studying the effects of the initial state
%interactions and are finally improving the matching with the QCD coefficient by
%improving the calculations in the spirit of Ref.\cite{Hambye:PRD}

%We demonstrated that the chiral theory enlarged by the introduction of vector
%mesons can eliminate quadratic divergences to $\mathcal{O}{(p^2/N_c)}$.
%The improved stability of $\rat$ is  encouraging to extent the calculation to the
%%factorizable terms and initial state interactions.
%We expect the changes to
%be small, but we plan to complete them and present them in a longer article.
%The extension of the method to the amplitudes $A_0$ and $A_2$ will
%involve additional operators $Q_1, Q_2, \dots$ with considerable
%increase in the computational work. It will be interesting,
%however, to check whether vector mesons make these amplitudes also
%more stable and whether the $\Delta I=1/2$ rule can be better
%reproduced in order to obtain a consistent prediction for $\rat$.

 Alternatively, it has been realized that a functional matching scheme introduced in
Ref.\cite{Wu00} can handle the problem and lead to an exact
matching. Two important matching conditions were obtained for
exactly matching chiral loop evaluation to QCD loop evaluation. A
set of chiral algebraic relations were demonstrated to hold at
chiral loop level. These chiral relations enable us to relate the
direct CP violation $\rat$ with $\Delta I=1/2$ rule. As a
consequence, the resulting predictions can lead to a consistent
explanation for both $\rat$ and $\Delta I=1/2$ rule.
%which is necessary for a satisfactory
%prediction on the direct CP violation $\rat$. The results are both
%renormalization scale and scheme independent.
Of particular, the results are no longer sensitive to the strange
quark mass. Therefore all the large uncertainties were
significantly reduced and it provided a more precise prediction
for the direct CP violation $\rat$.

%%%%%%%%%%%%%%%%%%%P2:  B phi-K%%%%%%%%%%%%%%%%%%%%%

We proceed to discuss CP violation in $B$ decays. With the
successful running of two $B$ factories in KEK and SLAC,  precise
measurements of the time-dependent CP asymmetries as well as the
directly CP asymmetries in rare $B$ decays become available.
%Among
%those interesting decay modes, the most important one, the CP
%asymmetry  of  $B \rightarrow J / \psi K_S$ has been successfully
%measured and agrees very well with the SM.
%and a very good agreement with the Standard Model (SM)
%prediction on  $\sin 2 \beta$  was found.
%
However, the recent Belle results on $\sin 2 \beta$ from $B
\rightarrow \phi K_S$, although with significant errors, have
indicated  that the value of $\sin 2 \beta$ from different decay
modes could be significantly different. The most recent
measurements on $\sin 2 \beta$ give $0.47 \pm 0.34^{+0.08}_{-0.06}
( \tmop{ Babar} )$ and $0.00\pm 0.23\pm0.05$ (Belle).
%Note that the prevous value given by Belle was
%$- 0.96 \pm 0.5^{+ 0.09}_{- 0.11}$ one year ago.
It is of course too early to draw any robust
conclusion. Nevertheless, it opens a possibility that large new
physics effects may show up in the $b \rightarrow s \bar s s$
processes.

There exist lots of possibilities for new physics. Here we would
like to consider a simple S2HDM4\cite{S2HDM4}. In this model,
there are new Yukawa interactions between Higgs bosons and heavy
fourth-generations quarks. Since in general the Yukawa interaction
is expected to be proportional to the coupled quark mass, the new
Yukawa couplings could be much stronger than that in the S2HDM
\cite{S2HDM} and SM4 . Unlike in the case of S2HDM, where the $b$
quark contribution to the QCD penguin diagram through neutral
Higgs boson loop is strongly suppressed by the small $b$ quark
mass, the same diagram with intermediate $b'$ quark may
significantly contribute to the related processes \cite{S2HDM4}.
This new feature only exists in this combined model, and is of
particular interest in studying the CP violation of $b\to s\bar s
s $ and other penguin dominant processes.

The Lagrangian for the S2HDM4  is \cite{S2HDM4b}
\begin{eqnarray*}
  \mathcal{L}_Y & = & \bar{\psi}_L Y^U_1  \widetilde{\phi_1} u_R +
  \bar{\psi}_L Y^D_1 \phi_1 d_R
  \nn
  &&+ \bar{\psi}_L Y^U_2  \widetilde{\phi_2} u_R +
  \bar{\psi}_L Y^D_2 \phi_2 d_R + H.c
\end{eqnarray*}
with the extended quark content of  $u_{L, R} = ( u, c, t, t'
)_{L, R}$ and $d_{L, R} = ( d, s, b, b' )_{L, R}$. The Yukawa
coupling matrices $Y^{U ( D )}_i$ are 4-dimensional matrices
accordingly. The two Higgs fields $\phi_1, \phi_2$ have vacuum
expectation values (VEV) of $v_1 e^{i \delta_1}$ and $v_2 e^{i
\delta_2}$ respectively, with $\sqrt{|v_1 |^2 + |v_1 |^2} = v =
246 \tmop{GeV} .$ The relative phase $\delta = \delta_1 -
\delta_2$ between two VEVs is physical and provides a new source
of CP violation.  In the mass eigenstates, the three physical
Higgs bosons are denoted by $H^0, A^0, \tmop{and} H^{\pm}$
respectively.  Due to the non-zero phase $\delta$,  all the Yukawa
couplings become complex numbers in the physical mass basis.
%even
%they are all real in the flavor basis. 
For simplicity, we assume
that the CKM matrix elements associating with $t'$, i.e. $V_{t'q}$
are negligible  and  only focus on the neural Higgs boson
contributions.

%%In the mass basis, the Yukawa interactions between neutral Higgs
%bosons  and quarks   have the following general form
%%\begin{eqnarray}
%$  \mathcal{L}_Y =  \eta^q_{ij}  \bar{q}_{iL} q_{jR} \phi + H.c.,$
%%\end{eqnarray}
%with $\phi = H^0$ or $A^0$. The Yukawa coupling $\eta^q_{\tmop{ij}}$ is
%usually parameterized as
%%\begin{eqnarray}
%$  \eta_{\tmop{ij}}^q=  \sqrt{m_{q_i} m_{q_j}} \xi_{q_i
%  q_j}/v$
%%\end{eqnarray}
%In the Chen-Sher ansartz {\cite{cheng:1987rs}} motivated by a
%Fritzsch type of Yukawa coupling matrix. the values of all
%$\xi_{q_i q_j}$s are of the same order of magnitude. However, from
%other textures of the coupling matrix the relations among
%$\xi_{q_i q_j}$s are different\cite{textures}. In the general
%case, they should be taken  as free parameters only to be
%determined or constrained by the experiments.

Note that the new contributions to QCD  and
electro(chromo)-magnetic operators depends on different parameter
sets. In the QCD penguin sector, the contribution depends on
$\xi_{b b'}^{*} \xi_{s b'}$ where in electro(chromo)-magnetic
sector it depends on both $\xi_{b' b}^{} \xi_{s b'}$ and $\xi_{b
b'}^{*} \xi_{s b'}$. It is convenient to define two weak phases
$\theta_1$ and $\theta_2 $ through
\begin{eqnarray*}
  &  & \xi_{b b'}^{*} \xi_{s b'} = | \xi_{b b'} \xi_{s b'
  } | e^{i \theta_1},  \xi_{b' b}^{} \xi_{s b'} = |
  \xi_{b' b}^{} \xi_{s b' } |e^{i \theta_2} .
%\nonumber\\
\end{eqnarray*}
Since in general $\xi_{b' b}^{}$ and $\xi_{b b'}^{*} $ are complex
numbers and $\xi_{b' b}^{} \neq \text{$\xi_{b b'}^{*} $}$, the two
phases are not necessary to be equivalent. The presence of two
rather than one independent phases is particular for this model,
which gives different contributions to the QCD penguin and
electro(chromo)-magnetic Wilson coefficients.  The interference
between them greatly enlarges the allowed parameter space.

Before making any predictions, one first needs to know how the new
parameters are constrained by other experiments. For the process
of concern, the most strict constraints comes from $b \rightarrow
s \bar{s} s$ processes such as $B \rightarrow X_s \gamma$ and
$B^0_s - \bar{B}^0_s$ mixing, etc.

The expression for $B \rightarrow X_s \gamma$ normalized to $B \rightarrow
X_c e \bar{\nu}_e$ reads
\begin{eqnarray}
  \frac{\tmop{Br} ( B \rightarrow X_s \gamma )}{\tmop{Br} ( B \rightarrow X_c
  e \bar{\nu}_e )} & = & \frac{6 |V_{\tmop{tb}} V_{\tmop{ts}}^{*} |^2
  \alpha_{\tmop{em}}}{\pi |V_{\tmop{cb}} |^2 f ( m_c / m_b )} |C_{7 \gamma} (
  \mu ) |^2
\nonumber\\
\end{eqnarray}
with $f ( z ) = 1 - 8 z^2 - 24 z^4 \ln z + 8 z^6 - z^8$ and
$\tmop{Br} ( B \rightarrow X_c e \bar{\nu}_e ) = 10.45\%$. The low
energy scale $\mu$ is set to be $ m_b$. Using the Wilson
coefficients at the  scale $m_W$ and running down to the $m_b$
scale through re-normalization group equations, we obtain the
predictions for  Br($B \rightarrow X_s \gamma$).  For simplicity,
we focus on the case in which the $b'$ contribution  dominates
through $H^0$ loop, namely, we push the masses of the charged
Higgs $H^{\pm}$ and the other pseudo-scalar boson $A^0$  to be
very high $( m_{H^{\pm}}, m_{A^0} > 500$ GeV) and ignore their
contributions. We take the following typical values of the
couplings
\begin{eqnarray}
  | \xi_{b b'} | &=& 50,  | \xi_{b' b} | = 0.8,  | \xi_{s b'} | = 0.8,
\nonumber\\
  m_{H^0} &=& m_b' = 200 \tmop{GeV}, \label{params}
\end{eqnarray}
%and give in Fig.1 of ref.\cite{X}
%Fig.\ref{BSG} the value of Br($B \rightarrow X_s
%\gamma$) as a function of $\theta_1$ with different values of
%$\theta_2$.
%
%%\tmfloat{h}{big}{figure}{\epsfig{file=BSG.eps}}{\label{BSG} The branching
%%ratio of $B \rightarrow X_s \gamma$ as functions of $\theta_2$ in the model of
%%S2HDM4 . The sold, dashed and dotted curves correspond to $\theta_1 = 1.5, 1.0
%%\tmop{and} 0.5$ respectively. Other parameters are taken from
%%Eq.(\ref{params}).}
%
%\begin{figure}[htb]
%\includegraphics[width=0.3\textwidth]{BSG.eps}
%\caption{The branching
%ratio of $B \rightarrow X_s \gamma$ as a functions of $\theta_2$ in the model of
%S2HDM4 . The solid, dashed and dotted curves correspond to $\theta_1 = 1.5, 1.0
%\tmop{and} 0.5$ respectively. Other parameters are taken from
%Eq.(\ref{params}).}
%\label{BSG}
%\end{figure}
%
%As indicated in the figure,
and found  two separated  ranges for parameters
$\theta_1$ and $\theta_2$ are allowed by the experiments.
\begin{eqnarray}
  &  & \text{$- 1.4 \lesssim \theta_2 \lesssim - 1.2$},  \text{ $0.4
  \lesssim \theta_2 \lesssim 0.7$}
\end{eqnarray}
for $0.5 \lesssim \theta_1\lesssim 1.5$
Note that we do not make a scan for the full parameter space, the
above obtained range are already enough for our purpose.

The other $b \rightarrow s \bar{s} s$ process which could impose
strong constraint is the mass difference of neutral $B_s^0 $
meson. The measurements from LEP give a lower bound of $\Delta
m_{B_s}>14.9 ps^{-1}$. In this model, the $b'$ contributes to
$\Delta m_{B_s}$ only through box-diagrams
%
%The box diagram
%contribution to $\Delta m_{B_s}$ is given by
%\begin{eqnarray}
%  \Delta m_{B_s} & = & \frac{G_F^2}{6 \pi^2} ( f_{B_s} \sqrt{B_{B_s}} )^2
%  m_{B_s} m_t^2 |V_{\tmop{ts}} |^2 \{ \eta_{\tmop{tt}} B^{\tmop{WW}} ( x_t )
%  \nn
%  &&+\frac{1}{4} \eta^{\tmop{HH}}_{\tmop{tt}} y_t | \xi_{tt} |^4 B^{\tmop{HH}}_V (
%  y_t ) \nonumber\\
%  &  & + 2 \eta_{\tmop{tt}}^{\tmop{HW}} y_t | \xi_{\tmop{tt}} |^2
%  B^{\tmop{HW}}_V ( y_t, y_w )
%  \nn
%  &&+ \frac{1}{4} \eta^{\tmop{HH}}_{\tmop{tt}} y' (
%  \frac{m_{b'} \sqrt{m_b m_s}}{2 V_{t b}^{} V_{t s}^{*} m_t^2} \xi_{b
%  b'}^{*} \xi_{s b'}^{} )^2 B^{\tmop{HH}}_V ( y' ) \}
%\end{eqnarray}
%where $G_F = 1.16 \times 10^{- 5} \tmop{GeV}^{- 2}$ is the Fermi
%constant. $f_{B_s}$ and $B_{B_s}$ are the decay constant and bag
%parameter for $B_s^0$. In the numerical calculations,  we take the
%value of $f_{B_s} \sqrt{B_{B_s}} =0.23$GeV. $\eta_{i j}$s are the
%QCD correction factors. of $B^{\tmop{HH}, \tmop{WW},
%\tmop{HW}}_{(V)}$ are loop integration functions. The mass ratios
%are defined as $y_t = m_t^2 / m_{H^{\pm}}^2, y_w = m_t^2 / m_W^2$
%and $y' = m_{b'}^2 / m_{H^0}^2$ respectively.  Note that in the
%mass difference of $B^0_s$ mesons, the contribution from S2HDM4
with the parameter $\xi_{b b'}^{*} \xi_{s b'}$.
%So,
%only the phase $\theta_1$ will present in the expression.
The numerical calculations show that the constraint is weak.

The neutron electric dipole moment (EDM) is expected to give
strong constraints on the new physics.
%In the SM, the neutron EDM is zero at even two loop
%level. The current experimental upper limit gives EDM$<1.1\times 10^{-25}
%ecm.$\cite{Hagiwara:2002fs}.  In general, the new physics contributes to the neutron EDM
%through one loop diagrams. In the presence of new scalars, additional
%significant contributions may arise, for example from the Weinberg gluonic operator
%\cite{Weinberg:1989dx} and also the two-loop Barr-Zee type diagrams \cite{Bjorken:1977vt,Barr:1990vd}
%etc.
%
However, all the above three type of mechanisms are not related to
$b\to s$ flavor-changing transitions and therefore will involve
different parameters.
%For the one-loop diagrams, the neutral EDM
%is mostly related to $\xi_{u(d)}$ and $\xi_{t(b')}$ through
%$u(d)-$quark EDM. For Weinberg three gluonic operator, the
%dominant contribution is from internal $b'$ loop. Thus it is
%related to $\xi_{b'b'}$.  Similarly, for two-loop Barr-Zee
%diagram, the $b'-$quark loop will play the most important role and
%the couplings involve only $\xi_{u(d)}, \xi_{b'b'}$ etc.
%
Thus the neutron EDM will impose strong constraints on other
parameters in this model and has less significance in current
studying of decay $B\to \phi K_S$. This is significantly different
from the S2HDM case in which the $t-$quark always domains the loop
contribution and the couplings $\xi_{tt}$ and $\xi_{bb}$ are
subjected to a strong constraint from neutron EDM.
Other constraints may  come from $K^0 - \overline{K^0}$ and $B^0_d
- \overline{B^0_d} $ mixings. But again those processes contain
additional free parameters such as the the Yukawa coupling of
$\xi_{b' d}$ and $\xi_{s b'}$, the constraints from those
processes are much weaker.

Now we are ready to discuss CP asymmetry in $B \rightarrow \phi
K_S$. The decay amplitude for $\bar{B} \rightarrow \phi
\bar{K^0}$reads

\begin{eqnarray}
  \mathcal{A}( \bar{B}^0_d \rightarrow \phi \bar{K}^0 ) & = & -
  \frac{G_F}{\sqrt{2}} V_{\tmop{ts}}^{*} V_{\tmop{tb}} ( a_3 + a_4 + a_5
  \nn
  &&-\frac{1}{2} ( a_7 + a_9 + a_{10} ) ) X,
\end{eqnarray}
with $X$ being a factor related to the hadronic matrix elements.
In the naive factorization approach $X = 2 f_{\phi} m_{\phi} (
\epsilon \cdot p_B ) F_1 ( m_{\phi} )$, where $\epsilon$, $p_B$,
$F_1$ are the polarization vector of $\phi$, the momentum of $B$
meson and form factor respectively. The coefficients $a_i $ are
known combinations of the Wilson coefficients.
%defined through the effective Wilson coefficients
%$C^{\tmop{eff}}_i s$ as follows
%\begin{eqnarray}
%  a_{2 i - 1} = C^{\tmop{eff}}_{2 i - 1} + \frac{1}{N_c} C^{\tmop{eff}}_{2 i}
%  & , & a_{2 i} = C^{\tmop{eff}}_{2 i} + \frac{1}{N_c} C^{\tmop{eff}}_{2 i -
%  1},
%\end{eqnarray}
Since the heavy particles such as $H^{\pm, 0}, A^0 \tmop{and} b'$
has been integrated out below the scale of $m_W$, the procedures
to obtain the effective Wilson coefficients $C^{eff}_i$ are
exactly the same as in SM.

\begin{figure}[htb]
\includegraphics[width=0.3\textwidth]{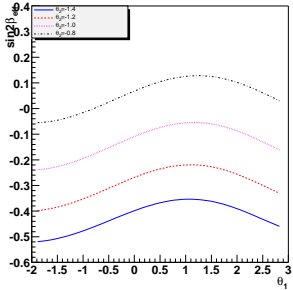}
\caption{The prediction for $\sin 2
\beta_{\tmop{eff}}$ as a functions of $\theta_1$ with different value of
$\theta_2$. The solid, dashed, dotted and dot-dashed curves corresponds to
$\theta_2 = - 1.4, - 1.2, - 1.0, - 0.8$ respectively.}
\label{BPHIK}                   %
\end{figure}

Using the above obtained parameters allowed by the current data,
the prediction for the time dependent CP asymmetry for $B
\rightarrow \phi K_S$  are shown in Fig.\ref{BPHIK}
%
%\tmfloat{h}{big}{figure}{
%
%\epsfig{file=BphiK-case2.eps,width=9cm}}{\label{BPHIK} The prediction for $\sin 2
%\beta_{\tmop{eff}}$ as functions of $\theta_1$ with diffrent value of
%$\theta_2$. The solid, dashed, dotted and dot-dashed curves corresponds to
%$\theta_2 = - 1.4, - 1.2, - 1.0, - 0.8$ respectively.}
%
In the figure, we give the value of $\sin 2 \beta_{\tmop{eff}}$ as
a function of $\theta_1$ with different values of
$\theta_2$=1.4,1.2,1.0 and 0.8. Comparing with the constraints
obtained from $B \rightarrow X_s \gamma$ and $B^0_s-\bar{B}^0_s$
mixings, one sees that in the allowed range of  $\text{$- 1.4 <
\theta_2 < - 1.2$} $ and $0.5 < \theta_1 < 1.5$, the predicted
$\sin 2 \beta_{\tmop{eff}}$ can reach $- 0.4$.
It is evident that the large negative value of $\sin2\beta_{eff}$
is a consequence of the interference effects between $\theta_1$
and $\theta_2$ and therefore is particular for this model.
%For
%zero value of $\theta_1$, there is no new phase in the QCD penguin
%sector. The allowed range for $\theta_2$ is $-1.0 \alt \theta_2
%\alt -0.8$\cite{Wu:2004kr}. Then, it follows from Fig.\ref{BPHIK},
%that in this range the predicted $\sin2\beta_{eff}$ is at around
%zero.  But for $\theta_1 \approx 0.5$, the allowed range for
%$\theta_2$ is changed into  $-1.4 \alt \theta_2 \alt -1.2$ and the
%predictions for $\sin2\beta_{eff}$ is much lower in the range of
%$(-0.4,-0.25)$.

In conclusion, we have briefly discussed
% the CP violation in SM,
%especially the more 
the precise consistent prediction for the direct
CP violation $\epsilon'/\epsilon$ in kaon decays and  the possible
large effects on CP violation in $B\to \phi K$ is discussed in a
model of S2HDM4 which contains both an additional Higgs doublet
and fourth generation quarks, since the fourth generation $b'$
quark is much heavier that $b$ quark, the Yukawa interactions
between neutral Higgs boson and $b'$ is greatly enhanced. This
results in significant modification to the QCD penguin diagrams.
The effective $\sin 2 \beta_{\tmop{eff}}$ in the decay $B
\rightarrow \phi K_S$ is predicted. We have shown that this model
can easily account for the possible large deviation of $\sin
2 \beta$ from it's SM value without conflicting with other experimental constraints.

\end{document}